\documentclass[aps,prl,reprint,nofootinbib,floatfix]{revtex4-2}
\usepackage[T1]{fontenc}
\usepackage{lmodern}
\usepackage[utf8]{inputenc}
\usepackage{amsmath,amssymb,bm,graphicx}
\usepackage{microtype}
\usepackage{needspace}
\usepackage{xcolor}
\usepackage[colorlinks=true,linkcolor=blue,citecolor=blue,urlcolor=black]{hyperref}

\newcommand{\manuscriptrefcolor}{blue}

\newcommand{\figref}[2][]{\textcolor{\manuscriptrefcolor}{Fig.~\ref{#2}#1}}

\newcommand{\bcite}[1]{\textcolor{\manuscriptrefcolor}{\cite{#1}}}
\newcommand{\smref}{\textcolor{\manuscriptrefcolor}{SM~\cite{supplement}}}
\newcommand{\smfirst}{\textcolor{\manuscriptrefcolor}{Supplemental Material (SM)~\cite{supplement}}}
\newcommand{\movref}[1]{\textcolor{\manuscriptrefcolor}{Movie~#1}}

\graphicspath{{figures/}}
\newcommand{\rr}{\bm r}
\newcommand{\pp}{\bm p}
\newcommand{\ttv}{\bm t}
\newcommand{\BB}{\bm B}
\newcommand{\Ree}{\operatorname{Re}}
\begin{document}
\title{Strain Waves Generate Compression and Trigger Buckling in Active Polymers}
\author{Wen-de Tian}
\email{tianwend@suda.edu.cn}
\author{ Kang Chen, Tianhui Zhang}
\affiliation{School of Physical Science and Technology, Soochow University, Suzhou 215006, China}

\begin{abstract}
Chemical reactions can sustain polymer motion without necessarily generating the compression required for buckling. Theory and simulations show that tension-dependent chemical transitions coupled to bond elasticity generate finite-wavelength traveling strain waves in a semiflexible polymer. In the straight chain at fixed axial span, the waves reduce mean active contraction while leaving mean elastic tension unchanged, generating compression that drives bending above a threshold. Releasing the mean load allows longitudinal waves to persist while transverse perturbations decay.
\end{abstract}

\maketitle

Chemical energy generates local forces in biological polymers, from cytoskeletal filaments to chromatin, yet their motion can involve many connected segments \bcite{prost2015,lizanaschwartz2024, sanchez2012}. In living nuclei, chromatin exhibits coordinated displacements over micrometer distances, and ATP depletion suppresses these spatial correlations \bcite{zidovska2013}. The relation between transcriptional activity and coherent motion depends on chromatin compaction \bcite{chu2024}. Models attribute such coherence to hydrodynamic cooperation of chromatin-bound motors \bcite{saintillan2018,eshghi2023} or to motor activity in cross-linked chains \bcite{liu2021}. However, such correlations alone do not reveal whether the chain carries a mean tensile or compressive force.

Mechanical load alters chemical transitions in molecular motors \bcite{visscher1999,block2003,laakso2008}. Motor activity coupled to elasticity can sustain filament beating \bcite{camalet1999,sanchez2011} and stress oscillations or propagating fronts in contractile materials \bcite{dierkes2014,banerjee2017,mondal2026, sakaue2026}.  Further, prescribed delayed local forces can coordinate polymer motion and induce looping through tension propagation \bcite{goychuk2024}. In disordered actomyosin assemblies, filament buckling can produce contraction \bcite{lenz2012,murrell2012}, while nonlinear elasticity more generally rectifies internal stresses into a net contractile stress \bcite{benoist2023}. This raises a question: Can autonomous longitudinal waves generate a mean compressive load in a straight chain with linear bonds and thereby trigger a separate bending instability?

Based on simulations and a continuum theory, we find that coupled chemical relaxation and elasticity can generate autonomous traveling waves.  These waves reduce mean active contraction, while linear bond elasticity leaves mean elastic tension unchanged at the fixed axial span. The  compression triggers a separate transverse instability and releasing the mean load preserves longitudinal waves but suppresses bending. Thus, nonlinear changes in chemical occupations connect internal motion to mean load and deformation.

\emph{Model and methods.---} The polymer is a three-dimensional  semiflexible chain with linear bonds, bending rigidity $\kappa$, and obeys the  overdamped dynamics [\figref[(a)]{fig:fig1}]. Bond $i$ carries elastic tension $T_i=k_b(\ell_i-b_0)$ and an active contractile force $a_i=\sum_{a=0}^2 f_a p_{ia}=\bm f^T\pp_i$. Here, $k_b$ is the spring stiffness, $\ell_i$ and $b_0$ are bond and rest lengths, and $\pp_i=(p_{i0},p_{i1},p_{i2})^T$ contains the occupations of the three states with forces $f_a$. The total bond force $S_i=T_i+a_i$ is tensile when positive and compressive when negative. The straight reference state has $S_i=0$, elastic tension $T_0$, and force-balanced bond length $\ell_0$. Define strain $\epsilon_i=(\ell_i-\ell_0)/b_0$ and contour wave number $q$ conjugate to $s_i=ib_0$.  Chemical transitions from state $a$ to state $b$ depend on the elastic tension through $k_{ab}(T;\chi)\propto \rho_k
\exp\!\left\{\gamma_{ab}\left[T_0+\chi(T-T_0)\right]\right\}$ , where the factor $\rho_k$ rescales all rates,  $\gamma_{ab}$ is a signed tension sensitivity, and $\chi$ changes the  tension sensitivity while preserving their values at $T_0$. This effective driven cycle violates detailed balance and can carry chemical currents at stationary occupations \bcite{julicher1997rmp}. Bond deformation changes occupations and the forces driving subsequent motion. We study periodic chains with either a fixed axial span or an adjustable cell length that relaxes the mean total bond force toward zero, and open chains with freely moving ends and no applied end forces. Details are given in \smfirst.

\begin{figure*}[!tbp]
\centering
\includegraphics[width=0.8\textwidth]{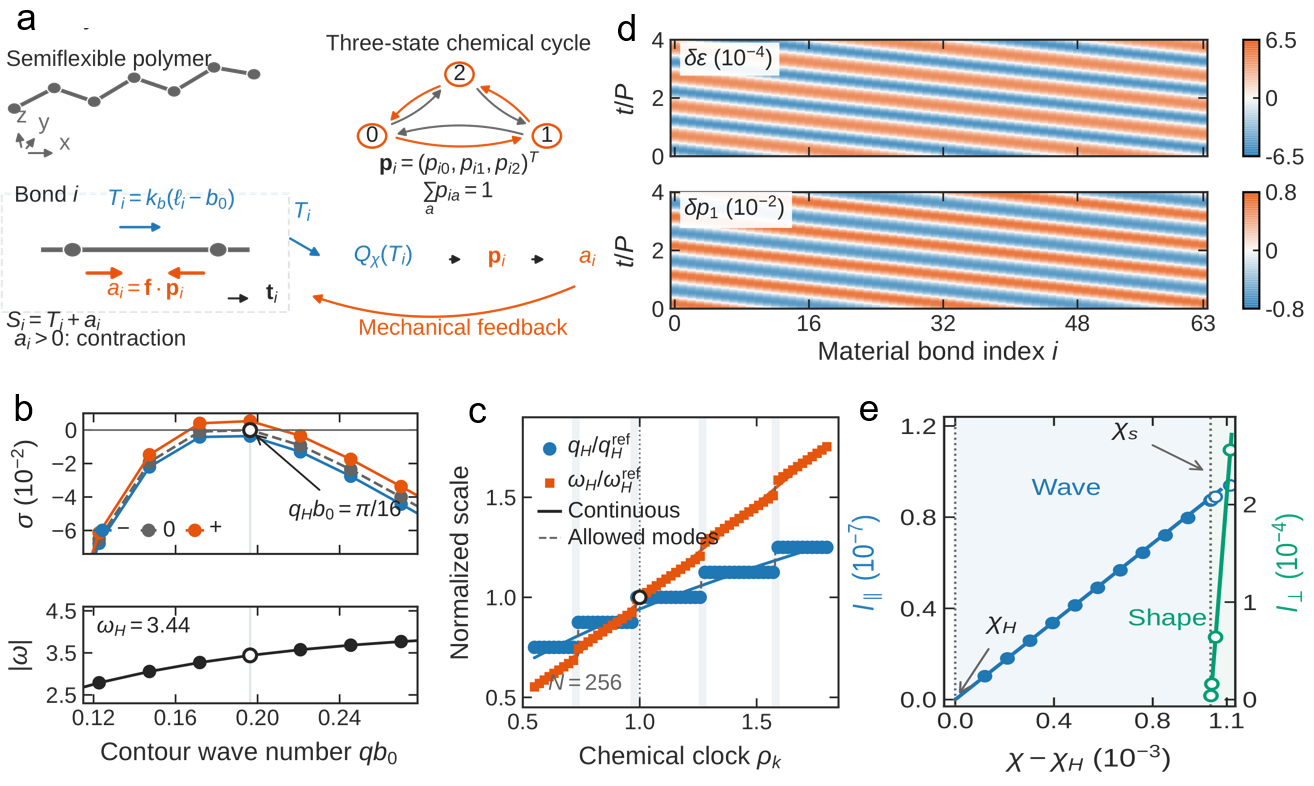}
\caption{Load-sensitive chemistry and traveling strain waves. (a) Bead-spring semiflexible polymer, bond forces, and three-state chemical cycle. (b) Longitudinal growth rate $\sigma=\Ree\lambda$ at $\chi_H-0.002$ (blue), $\chi_H$ (gray), and $\chi_H+0.003$ (orange), and oscillation frequency $|\omega(q)|$ at $\chi_H$ (lower), versus wave number $q$; $N=256$. Open circles mark $q_H$. (c) Onset wave number (circles) and frequency (squares), normalized by $q_H^{\rm ref}$ and $\omega_H^{\rm ref}$ at $\rho_k=1$, versus the common chemical-rate factor $\rho_k$. Solid/dashed lines: continuous/allowed-mode predictions; symbols: $N=256$. Shading marks intervals containing switches between allowed modes. (d) Strain $\delta\epsilon_i=\epsilon_i-\bar\epsilon_i$ and state-1 occupation $\delta p_{i1}=p_{i1}-\bar p_{i1}$ versus bond index $i$ and time $t/P$; $N=64$, $\chi=\chi_s$. Overbars denote period means, and $P$ is the wave period. (e) Longitudinal intensity $I_\parallel$ (blue) and transverse intensity $I_\perp$ (green) versus $\chi-\chi_H$; $N=64$. Filled/solid and open/dashed blue branches are transversely stable/unstable; green circles: curved states. Lines: amplitude theory; dotted lines: longitudinal onset $\chi_H$ and bending threshold $\chi_s$. Unless indicated, $\kappa=1.059$ and $b_0=\zeta=1$.}
\label{fig:fig1}
\end{figure*}

\emph{Autonomous waves.---} We first examine the longitudinal stability of a straight chain at fixed axial span. For the reference kinetics at \(N=256\), increasing the tension sensitivity \(\chi\) produces a finite-wave-number Hopf instability at \(\chi_H\simeq0.935\). The first unstable mode has a spatial period of 32 bonds [\figref[(b)]{fig:fig1}]. The leading eigenvalue \(\lambda(q)\) of the linearized dynamics gives the growth rate \(\sigma(q)=\operatorname{Re}\lambda(q)\) and oscillation frequency \(|\omega(q)|=|\operatorname{Im}\lambda(q)|\). At fixed tension, perturbations in the chemical occupations decay. When bonds deform, however, the occupations respond with a finite delay, producing a phase difference between elastic tension and active contraction. For the reference kinetics, the delayed changes in active contraction amplify intermediate-wavelength oscillations, while long-wavelength and bond-scale perturbations decay.

Rescaling all chemical rates by $\rho_k$ preserves stationary occupations at fixed tension but changes the onset frequency and wavelength [\figref[(c)]{fig:fig1}]. In the long-wavelength continuum limit, the onset frequency scales as $\rho_k$ and the wavelength as $\rho_k^{-1/2}$. Because the elastic relaxation rate grows as $q^2$, faster chemistry couples to faster mechanical relaxation at shorter wavelengths. Finite chains select among discrete wave numbers, so the selected wave number changes in steps as \(\rho_k\) is varied. 
The two propagation directions are equivalent at onset. Near onset, each traveling component suppresses its counterpropagating counterpart more strongly than it limits its own growth, selecting one propagation direction (End Matter). The  traveling wave [\figref[(d)]{fig:fig1}] remains straight over a finite interval above $\chi_H$, before becoming unstable to bending at $\chi_s$ [\figref[(e)]{fig:fig1}, \movref{1}].

\emph{Load becomes compressive.---}
We next examine how longitudinal waves alter the mean bond force. For a straight periodic chain at fixed axial span, linear elasticity fixes the mean elastic tension at $\bar T=T_0$, where overbars denote averages over bonds and one wave period. During a wave, oscillations in tension and chemical occupations change the average transition fluxes at second order in wave amplitude, shifting the mean occupations. The mean occupation of the strongly contractile state 1 decreases, while that of the weakly contractile state 2 increases [\figref[(a)]{fig:fig2}]. State 0 carries no active force. The loss of contraction from state 1 exceeds the gain from state 2, reducing $\bar a$. Active contraction then no longer balances the outward elastic force of the precompressed bonds, leaving $\bar S=T_0+\bar a<0$ while the chain is still straight. Near onset, $-\bar S=\Gamma I_\parallel$, where $I_\parallel$ is the sum of the squared Fourier amplitudes of the two counterpropagating strain modes. The reference kinetics gives $\Gamma>0$ [\figref[(b)]{fig:fig2}], although its sign depends on the chemical network and state forces.

To distinguish the role of chemical relaxation in wave generation from its role in compression, we compare the dynamic occupations with the stationary occupations evaluated at the instantaneous tension along the same prescribed weak tension waveform. Both responses reduce the mean active force. At the reference Hopf point, finite relaxation reduces compression per unit wave intensity by about $24\%$ relative to the instantaneous response (\smref). This means that finite chemical relaxation sustains the autonomous wave in this model and modifies its mean load, whereas compression under a prescribed tension cycle does not require occupation delay.

\begin{figure}[!tbp]
\centering
\includegraphics[width=0.95\columnwidth]{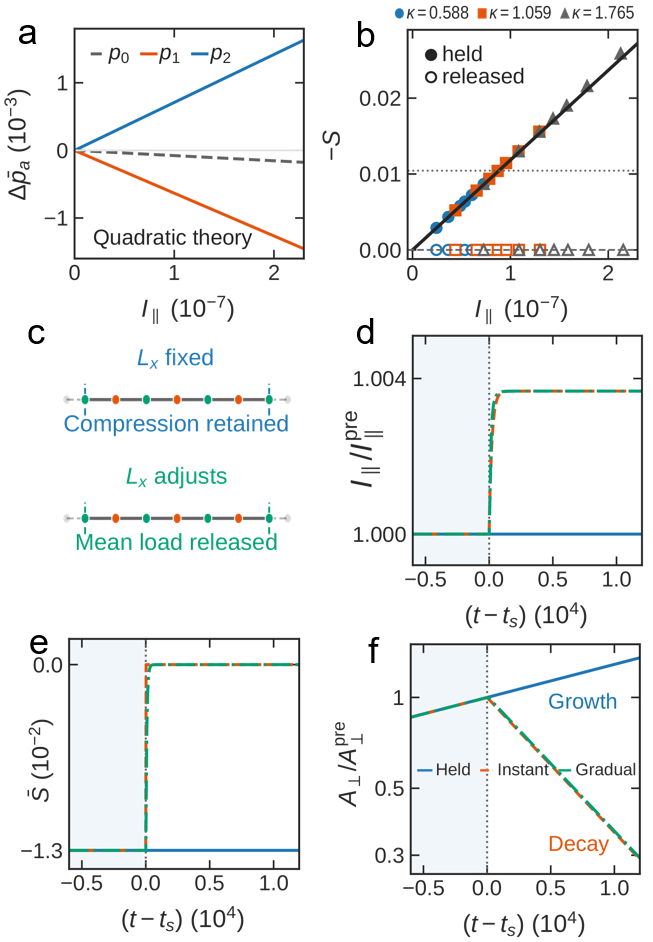}
\caption{Load release preserves longitudinal waves and suppresses bending. (a) Mean occupation shifts versus longitudinal intensity $I_\parallel$: states 0 (gray dashed), 1 (strongly contractile, orange), and 2 (weakly contractile, blue). (b) Mean compression $-\bar S$ versus $I_\parallel$ for fixed-span (filled) and released (open) waves at the indicated $\kappa$. Solid line: $-\bar S=\Gamma I_\parallel$; dotted line: first bending threshold at $N=64$, $\kappa=1.059$. (c) Fixed and adjustable periodic spans; pale beads are periodic copies. (d) Longitudinal intensity $I_\parallel/I_\parallel^{\rm pre}$, (e) mean bond force $\bar S$, and (f) transverse Fourier amplitude $A_\perp/A_\perp^{\rm pre}$ versus time for fixed-span (blue solid), immediate-release (orange dashed), and gradual-release (green dash-dotted) evolution from the same wave. Superscript ``pre'': value before release; dotted lines: release time $t_s$. For (d--f), $N=64$, $\chi=0.936$, $\kappa=1.059$, and gradual-release time $\tau_c=200$.}
\label{fig:fig2}
\end{figure}

To test whether mean compression drives bending, we initialize two periodic chains from the same longitudinal traveling-wave state with the same small transverse perturbation [\figref[(c)]{fig:fig2}]. The axial span is fixed in one and adjusted in the other so that the spatially averaged total bond force relaxes toward zero. Span adjustment also changes elastic tension and chemical rates. For the state shown in \figref[(d)]{fig:fig2}, immediate load release changes the longitudinal wave intensity by only about $0.4\%$, but turns transverse growth into decay [\figref[(e,f)]{fig:fig2}]. Gradual release likewise turns growth into decay as the mean compression weakens [\figref[(e,f)]{fig:fig2}, \movref{2}]. 

\emph{Compression controls bending.---}
We next examine the transverse stability of the straight traveling-wave state. Uniform mean compression produces no net force on each bead of a straight periodic chain. A small transverse deflection tilts adjacent bonds, allowing the compressive forces to amplify the deflection, while bending elasticity opposes it. When transverse deflections evolve over many longitudinal wave periods, period averaging gives the growth rate of transverse mode $m$, $\lambda_{\perp,m}\simeq-\bar S K_m/(\zeta\ell_0)-\kappa K_m^2/(\zeta\ell_0^2)$, where $K_m=4\sin^2(\pi m/N)$ for an $N$-bead periodic chain and $\zeta$ is the bead drag coefficient. Compression makes the first term positive, whereas bending elasticity suppresses short-wavelength distortions more strongly. This balance accounts for the transverse growth rates over the examined chain lengths, rigidities, and constraints [\figref[(a)]{fig:fig3}]. The longest nonuniform transverse mode, $m=1$, becomes unstable first, when $-\bar S>\kappa K_1/\ell_0$. For $\Gamma>0$, $\chi_s-\chi_H\propto N^{-2}$ for long chains containing integer periods of the same longitudinal wave [\figref[(b)]{fig:fig3}]. When the mean load relaxes to zero, the compressive contribution vanishes and bending elasticity damps small transverse deflections. At fixed parameters, the slowest nonuniform mode then has $\lambda_{\perp,1}\propto-N^{-4}$ [\figref[(c)]{fig:fig3}].

At fixed axial span, bending increases the chain's contour length and mean bond extension, raising the mean elastic tension. The  tension changes alter chemical occupations and the longitudinal waveform. On the weakly curved branch, the mean active contraction also increases. Relative to the straight traveling wave at the same $\chi$, both $\Delta\bar T$ and $\Delta\bar a$ are positive, so $\Delta\bar S=\Delta\bar T+\Delta\bar a>0$ and the mean compression weakens [\figref[(d)]{fig:fig3}]. This feedback limits further bending [\movref{3}]. The saturated state retains a compressive mean force close to $-\kappa K_1/\ell_0$, where the period-averaged effects of compression and bending elasticity balance. Near the bending threshold, the transverse intensity $I_\perp$, proportional to the squared bending amplitude, increases linearly with $\chi-\chi_s$, as predicted by the cubic amplitude theory [\figref[(e)]{fig:fig3}]. The cubic saturation coefficient estimated from the mean-force change is close to the full result; the full calculation also includes longitudinal waveform changes and second harmonics (End Matter). The saturated chain bends in both transverse directions, producing a nearly circular $y$--$z$ projection [\figref[(f)]{fig:fig3}].

\begin{figure}[!tbp]
\centering
\includegraphics[width=0.95\columnwidth]{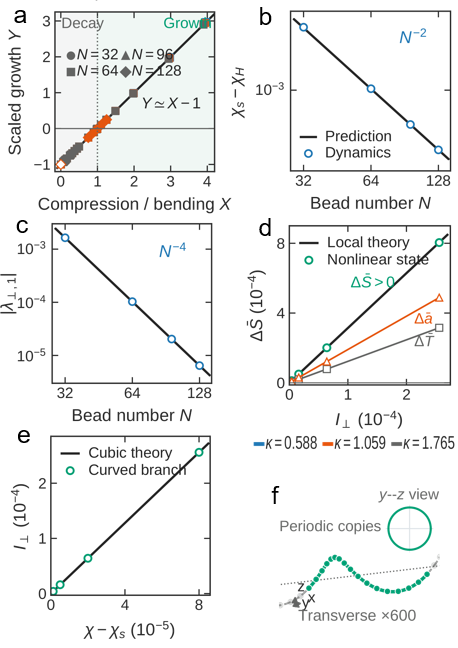}
\caption{Mean compression controls the onset and saturation of bending. (a) Scaled growth rate $Y=\zeta\ell_0^2\lambda_{\perp,m}/(\kappa K_m^2)$ versus scaled compression $X=-\bar S\ell_0/(\kappa K_m)$, where $K_m=4\sin^2(\pi m/N)$. Symbol shapes/colors denote $N$/$\kappa$; filled/open symbols: fixed-span/released states. Line: $Y=X-1$. (b) Threshold separation $\chi_s-\chi_H$ and (c) first-mode decay rate $|\lambda_{\perp,1}|$ after mean-load release versus bead number $N$. Circles: full dynamics; lines: discrete-chain predictions approaching $N^{-2}$ and $N^{-4}$, respectively. (d) Mean-force changes relative to the straight wave at the same $\chi$, versus transverse intensity $I_\perp$: total $\Delta\bar S$ (green circles), active $\Delta\bar a$ (orange triangles), and elastic $\Delta\bar T$ (gray squares); lines: theory. (e) Transverse intensity $I_\perp$ versus $\chi-\chi_s$ on the curved branch (circles), with cubic amplitude theory (line). (f) Near-threshold curved state; transverse magnification $600$, pale beads: periodic copies; inset: $y$--$z$ view. For (b--f), $\kappa=1.059$; for (d--f), $N=64$.}
\label{fig:fig3}
\end{figure}

\begin{figure}[!tbp]
\centering
\includegraphics[width=0.95\columnwidth]{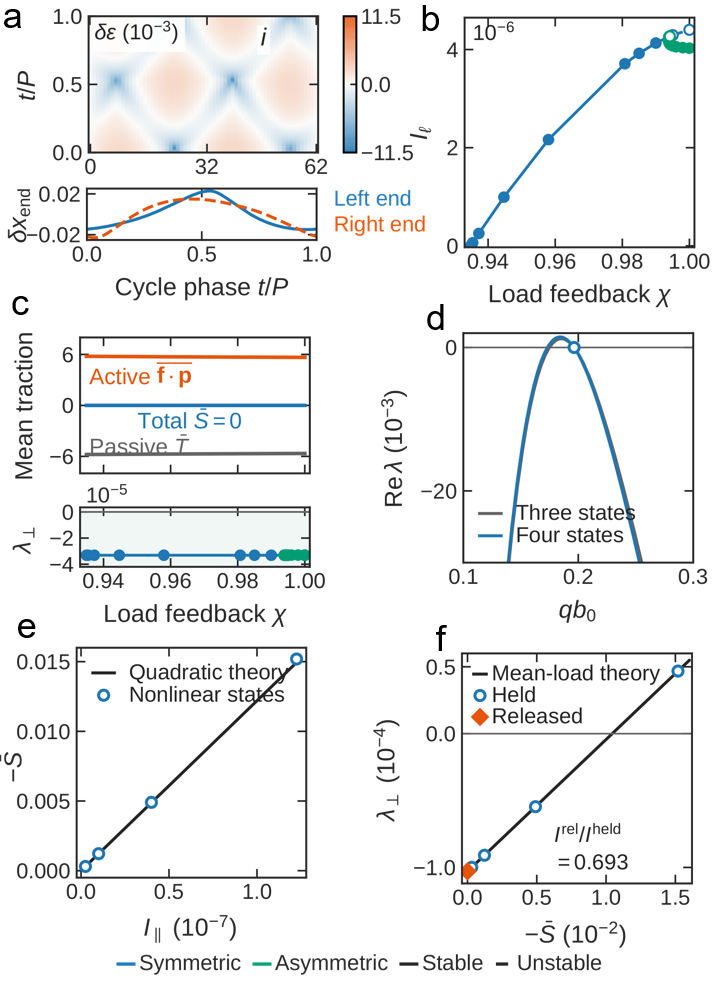}
\caption{Free-end oscillations and a four-state chemical cycle. (a) Period-mean-subtracted strain $\delta\epsilon$ versus bond index $i$ and time $t/P$ (upper), and endpoint displacements $\delta x_{\rm end}$ versus $t/P$ (lower), at $\chi=0.990$; $P$ is the period. Blue solid/orange dashed: left/right ends. (b) Mean-square strain $I_\ell$ versus $\chi$ on free-end branches. (c) Mean active (orange), elastic (gray), and total (blue) forces (upper), and largest transverse growth rate $\lambda_\perp$ excluding rigid translations and rotations (lower), versus $\chi$. Blue/green in (b) and lower (c): symmetric/asymmetric waveforms. Symmetric branches obey $\ell_i(t)=\ell_{N-2-i}(t+P/2)$. In (b), solid/filled and dashed/open marks: longitudinal stability/instability. (d) Three-state (gray) and four-state (blue) growth rates versus $q$ at their respective onsets; open circle: critical allowed mode. (e) Four-state mean compression versus longitudinal intensity $I_\parallel$: nonlinear waves (circles) and weak-wave theory (line). (f) First-mode transverse growth rate versus mean compression $-\bar S$ for fixed-span (circles) and released (diamond) waves; line: mean-load prediction. The released wave and the most compressed fixed-span wave have the same $\chi$, with $I_\parallel^{\rm released}/I_\parallel^{\rm held}=0.693$. $N=64$, $\kappa=1.059$; (d--f): periodic chains with added-channel strength $\eta=0.300$ (\smref).}
\label{fig:fig4}
\end{figure}

\emph{Free ends and a four-state chemical cycle.---}
We next examine an open chain that adjusts its extension through freely moving endpoints, with no applied end forces. Bond lengths and endpoint positions oscillate periodically [\figref[(a)]{fig:fig4}, \movref{4}], while the mean extension is free to adjust. The mean-square strain $I_\ell=(N-1)^{-1}\sum_i\overline{\epsilon_i^2}$ includes squared mean strain offsets and oscillatory contributions, whereas $I_\parallel$ measures only the selected traveling-wave modes [\figref[(b)]{fig:fig4}]. For a straight free-end state that is periodic in time, period-averaged force balance and the absence of end forces require $\bar S_i=0$ in every bond, although instantaneous internal forces continue to oscillate [\figref[(b)]{fig:boundary}]. A separate Floquet analysis gives a negative largest nonneutral transverse growth rate on the examined branches, confirming that small transverse perturbations decay [\figref[(c)]{fig:fig4}]. These straight branches differ in the symmetry of their longitudinal waveforms; some are longitudinally unstable despite being transversely stable.

To test whether wave-induced compression persists beyond the three-state model, we add a fourth force-generating state, introducing an additional chemical relaxation process (\smref). The extended model still undergoes an oscillatory instability at a finite wave number [\figref[(d)]{fig:fig4}]. The  traveling waves reduce the mean active contraction and generate compression at fixed axial span [\figref[(e)]{fig:fig4}]. At the same $\chi$, we compute traveling-wave states separately under fixed-span and mean-load-release conditions. Small transverse perturbations grow in the fixed-span state with the largest compression and decay in the corresponding released state [\figref[(f)]{fig:fig4}]. Traveling-wave selection and compression also persist under the tested small changes in rate parameters and state forces that preserve oscillatory onset (\smref).

\emph{Discussion.---}
Tangential motor loads can directly destabilize a straight filament \bcite{decanio2017}. Here autonomous longitudinal waves redistribute chemical occupations and, for the cycle studied, reduce the mean active contraction. For the straight chain at fixed axial span, linear bond elasticity keeps the mean elastic force unchanged, leaving compression that drives a separate transverse instability. Sustained waves therefore do not by themselves imply bending: the mechanical constraint must retain sufficient mean compression.

Chromatin cross-links and nuclear-envelope attachments modify the collective response to motor forces \bcite{lewis2026,schreiner2015,kalukula2022}. Applying the wave-induced compression mechanism to chromatin would require tension-sensitive chemical transitions and attachments that hold axial separation approximately fixed for longer than the bending growth time. Tracking longitudinal wave intensity, mean axial force, and transverse perturbations before and after attachment release would test this possibility. If waves persist while compression falls below the bending threshold and transverse growth turns into decay, the response would support wave-generated compression as a mechanism for bending.

\begin{acknowledgments}
This work was supported by the National Natural Science Foundation of China (Grant No. 22573069).
\end{acknowledgments}

\clearpage

\renewcommand{\theHfigure}{E\arabic{figure}}

\section*{End Matter}
\renewcommand{\thefigure}{E\arabic{figure}}
\setcounter{figure}{0}

\paragraph*{Dynamics and mechanical constraints.}
Let $\bm d_i=\rr_{i+1}-\rr_i$, $\ell_i=|\bm d_i|$, and $\ttv_i=\bm d_i/\ell_i$. The passive energy and deterministic equations are $U=\frac{k_b}{2}\sum_i(\ell_i-b_0)^2 +\kappa\sum_i(1-\ttv_i\cdot\ttv_{i-1}),$ $\zeta\dot\rr_i=S_i\ttv_i-S_{i-1}\ttv_{i-1} +\bm F_i^{\rm bend},$ $\dot\pp_i=Q(T_i;\chi)\pp_i,\qquad \bm1^T\pp_i=1.$ Here $\bm F_i^{\rm bend}$ is minus the gradient of the bending energy, $Q_{ba}=k_{ab}$ for $b\ne a$, and each column of $Q$ sums to zero. Only the elastic tension $T_i$, rather than the total force $S_i$, enters the transition rates in the main text. The notation $Q_\chi(T_i)$ in \figref[(a)]{fig:fig1} means $Q(T_i;\chi)$. The driven cycle violates detailed balance even when its occupations are stationary. The main calculations omit excluded-volume forces; the additional fluctuating-chain calculations use a Weeks--Chandler--Andersen potential~\bcite{wca}. At $k_b=250$, $b_0=\zeta=1$, the rounded reference values are $T_0=-5.779$, $\ell_0=0.977$, and $\pp^*=(0.310,0.254,0.436)^T$. The state forces are $(f_0,f_1,f_2)=(0,20.312,1.429)$, with reference active force $a^*(T_0)=\bm f^T\pp^*=-T_0$ and $S_0=0$. Numerical units are not calibrated to biological time or force.

The fixed periodic span obeys $\rr_{i+N}=\rr_i+L_x\bm e_x$, with $L_x=N\ell_0$. Mean-load release rescales the axial coordinates and $L_x$, immediately or with response time $\tau_c=200$, until the spatial mean of $S_i$ vanishes. The chemical law and initial occupations are retained, while the waveform evolves freely after release. An open $N$-bead chain has $N-1$ bonds and freely moving endpoints (\figref[]{fig:boundary}). Overbars denote averages over one oscillation period, with an additional bond average for quantities without a bond index.

\begin{figure*}[t]
\centering\includegraphics[width=0.8\textwidth]{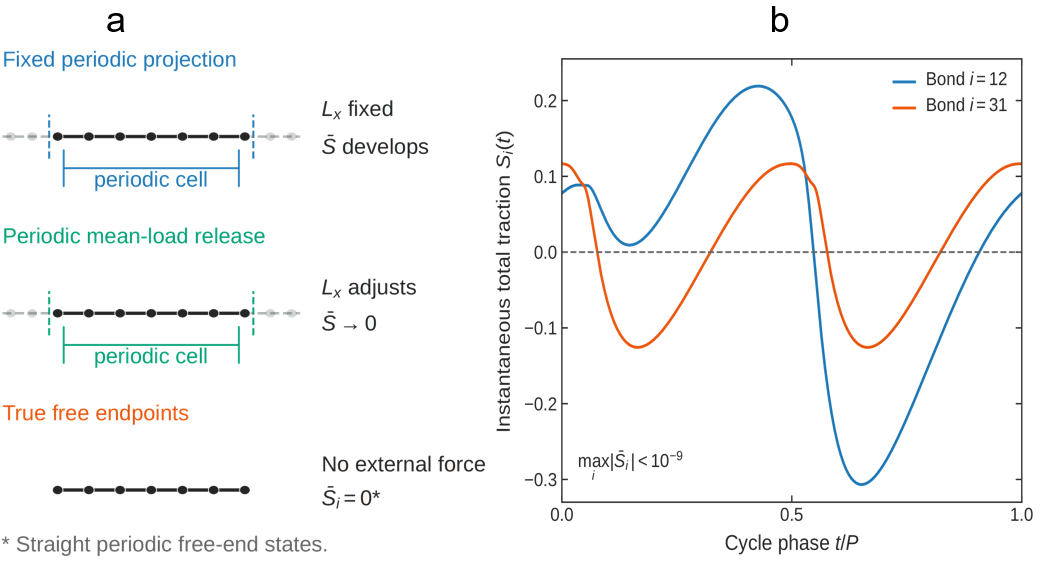}
\caption{Mechanical constraints and oscillating internal forces. (a) Periodic chains with fixed and adjustable spans $L_x$, and an open chain with freely moving endpoints. Pale beads are periodic copies. Straight, time-periodic free-end states have zero cycle-averaged force in each bond, $\bar S_i=0$. (b) Instantaneous total bond forces $S_i(t)=T_i(t)+\bm f^T\pp_i(t)$ versus time at bonds $i=12$ (blue) and $31$ (orange), for $N=64$, $\chi=0.994$. Forces oscillate about zero mean; the dashed line marks zero force. Time is normalized by the period $P$; $b_0=\zeta=1$.}
\label{fig:boundary}
\end{figure*}

\paragraph*{Longitudinal response and mean compression.}
In independent chemical coordinates $\bm z$, write $\dot{\bm z}=\bm F(\bm z,T)$ and $a=a(\bm z)$. Define $J=\partial_{\bm z}\bm F$, $\bm g=\partial_T\bm F$, and $\bm h=\nabla_{\bm z}a$ at the reference state and chosen $\chi$. Eliminating occupation perturbations gives $\lambda+c(q)[1+\mathcal R(\lambda)]=0,$ $\mathcal R(\lambda)=\bm h^T(\lambda I-J)^{-1}\bm g,$ where $c(q)=4k_b\sin^2(qb_0/2)/\zeta$, and $q$ is measured in the material coordinate $s_i=ib_0$. At $N=256$ and $\rho_k=1$, the reference onset has $q_Hb_0=\pi/16$, $\omega_H=3.440$, and $\chi_H=0.935$. Continuous-wave-number scaling applies while the optimum remains accessible; finite chains select allowed discrete modes.

Define $\epsilon_i=(\ell_i-\ell_0)/b_0$. Its leading wave component is $A_+e^{i(q_Hib_0+\omega_Ht)}+A_-e^{i(-q_Hib_0+\omega_Ht)}+\mathrm{c.c.}$, with $I_\parallel=|A_+|^2+|A_-|^2$. Near onset, the amplitude magnitudes obey $d|A_\pm|/dt=[\sigma_\chi(\chi-\chi_H)-g_s|A_\pm|^2-g_c|A_\mp|^2]|A_\pm|$. The growth-rate slope is $\sigma_\chi=\left.\partial_\chi\Ree\lambda(q_H)\right|_{\chi_H}=1.789$, and the self- and cross-saturation coefficients are $g_s=20886.048$ and $g_c=54203.292$. Thus $g_c>g_s>0$ selects a traveling wave, with $I_\parallel\simeq\sigma_\chi(\chi-\chi_H)/g_s$ near onset.

For the straight fixed-span wave, $\theta=T-T_0$ has zero mean. The second-order occupation shift satisfies $Q_0\Delta\bar\pp=-Q_T\langle\theta\delta\pp^{(1)}\rangle -\tfrac12Q_{TT}\pp^*\langle\theta^2\rangle,$ with $\bm1^T\Delta\bar\pp=0$. Here $Q_0=Q(T_0;\chi_H)$, $Q_T=\left.\partial_TQ\right|_{T_0,\chi_H}$, and $Q_{TT}=\left.\partial_T^2Q\right|_{T_0,\chi_H}$; $\delta\pp^{(1)}$ is the first-order occupation response, and angle brackets denote averages over bonds and one period. The first term captures the correlation between tension and the occupation response; the second captures the nonlinear tension dependence of the transition rates. Together they give $\Gamma=118346.272$ in $-\bar S\simeq\Gamma I_\parallel$. The sign follows from this response and the state forces; a different chemical network need not give $\Gamma>0$.

\paragraph*{Bending threshold and saturation.}
With $K_m=4\sin^2(\pi m/N)$, define the scaled load $X=-\bar S\ell_0/(\kappa K_m)$ and growth rate $Y=\zeta\ell_0^2\lambda_{\perp,m}/(\kappa K_m^2)$. The near-onset averaging law in the main text gives $Y=X-1$ and $\chi_s-\chi_H\simeq\frac{g_s\kappa K_1}{\sigma_\chi\Gamma\ell_0}.$ At $N=64$, the leading estimate is $1.030\times10^{-3}$; full continuous-time stability gives $1.034\times10^{-3}$. The $N^{-2}$ threshold separation and $N^{-4}$ released decay scaling concern commensurate chains carrying the same primary wave.

In the frame moving with the wave, $\BB$ is the two-component complex amplitude of the critical $m=1$ transverse displacement mode. Through cubic order, $\dot\BB=(a_s\delta+i\Omega_s)\BB -\mathcal G_1(\BB^\dagger\BB)\BB -\mathcal G_2(\BB\cdot\BB)\BB^*,$ where $\delta=\chi-\chi_s$ and $G_j=\Ree\mathcal G_j$. The coefficients include direct cubic terms, changes in the mean state and longitudinal wave, and second harmonics. At $N=64$, $a_s=9.912\times10^{-2}$, $G_1=6.197\times10^{-2}$, and $G_2=1.553\times10^{-6}$. Define $I_\perp=2\BB^\dagger\BB$, where $\BB$ has displacement units and $I_\perp$ has squared-length units. For the single-mode Fourier probe in \figref[(f)]{fig:fig2}, $A_\perp^2=2I_\perp$. In the linear regime, the amplitude grows at $\lambda_\perp$, while the intensity grows at $2\lambda_\perp$. For a circular transverse profile, $\BB\cdot\BB=0$ and $I_\perp\simeq2a_s\delta/G_1$; a planar profile instead involves $G_1+G_2$. For the weakly curved branch in \figref[(d)]{fig:fig3}, the mean-force change is measured relative to the straight wave at the same $\chi$. Its local slope $\Delta\bar S/I_\perp\simeq3.143$ gives $G_1^{\rm load}=2K_1(\Delta\bar S/I_\perp)/(\zeta\ell_0)\simeq6.197\times10^{-2}$ through the averaged transverse-growth law, close to the complete $G_1$. This relation connects the cubic saturation coefficient to the total mean-force response but does not separate the individual cubic contributions. Since $G_2>0$ but $G_2\ll G_1$, the cubic theory predicts slow ellipticity relaxation asymptotically close to onset; the theory does not select an absolute spatial orientation or determine shapes far from onset. Nonlinear saturation here is established for the three-state cycle.

\paragraph*{Free-end balance and stability.}
For a straight, time-periodic open chain, $\zeta\dot x_i=S_i-S_{i-1}$, with $S_{-1}=S_{N-1}=0$ at the free ends. Period averaging and zero external endpoint forces give $\bar S_i=0$ in every bond, while instantaneous forces remain finite [\figref[(b)]{fig:boundary}]. Internal force pairs fix the center of mass under local drag, so periodic bond lengths also imply periodic bead positions. The longitudinal observable is $I_\ell=(N-1)^{-1}\sum_i\overline{\epsilon_i^2}$, including mean-extension changes. A Floquet multiplier $\Lambda$ over period $P$ has growth rate $P^{-1}\ln|\Lambda|$. Time-shift, translation, and rotation neutral modes are excluded from the relevant tests; longitudinal and transverse stability are evaluated separately. Zero mean bond force does not by itself establish transverse stability; transverse decay is determined separately from the Floquet spectrum.

\paragraph*{Instantaneous occupation response.}
Let $a^*(T)=\bm f^T\pp^*(T)$, where $Q(T;\chi_H)\pp^*(T)=0$ and $\bm1^T\pp^*(T)=1$. Its derivatives obey $Q_0\pp_T^*=-Q_T\pp^*,$ $Q_0\pp_{TT}^*=-2Q_T\pp_T^*-Q_{TT}\pp^*.$ Both derivatives have zero component sum. For the same weak tension harmonic, $\Gamma_{\rm ad}=-(k_bb_0)^2\bm f^T\pp_{TT}^*=155917.314$, whereas finite relaxation gives $\Gamma=118346.272$, a $24.1\%$ reduction at the reference onset. Instantaneous-occupation eigenvalues are $\lambda=-c(q)[1+\partial_Ta^*]$, with $\partial_Ta^*=1.579$; every nonuniform longitudinal mode decays at this reference state. Thus instantaneous occupations can generate compression under a supplied tension cycle, while finite relaxation sustains the autonomous wave. Further kinetic parameters, convergence checks, and the four-state extension are given in \smref.


\begin{thebibliography}{31}
\bibitem{supplement} See SM at [URL will be inserted by the publisher] for kinetic parameters, response derivations, nonlinear branches, stability calculations, noise protocols, convergence checks, numerical data, and analysis code.
\end{thebibliography}
\end{document}